\documentclass[fleqn,usenatbib,letters]{mnras}

\usepackage{newtxtext,newtxmath}
\usepackage[T1]{fontenc}

\DeclareRobustCommand{\VAN}[3]{#2}
\let\VANthebibliography\thebibliography
\def\thebibliography{\DeclareRobustCommand{\VAN}[3]{##3}\VANthebibliography}

\usepackage{graphicx}	% Including figure files
\usepackage{amsmath}	% Advanced maths commands
\usepackage{lipsum}
\usepackage{algorithm}
\usepackage{algorithmic}
\usepackage{bm}

\usepackage{xcolor}
\renewcommand{\v}[1]{\mathbf{#1}}
\newcommand{\vg}[1]{\boldsymbol #1}
\newcommand{\T}{\ensuremath{^\top}}
\newcommand{\solarmass}{\ensuremath{M_\odot}}
\defcitealias{behrooziComprehensiveAnalysisUncertainties2010}{B10}
\defcitealias{fischbacherGalSBIPhenomenologicalGalaxy2024}{F25}

\newenvironment{talign*}
 {\let\displaystyle\textstyle\csname align*\endcsname}
 {\endalign}

\title{SHAM-OT: Rapid Subhalo Abundance Matching with Optimal Transport}

\author[Fischbacher \& Kacprzak et al.]{
Silvan Fischbacher,$^{1}$\thanks{Equal contribution. E-mail: silvanf@phys.ethz.ch},
Tomasz Kacprzak$^{2}$\thanks{Equal contribution. E-mail: tomasz.kacprzak@psi.ch},
Luis Fernando Machado Poletti Valle,$^{1}$
Alexandre Refregier$^{1}$
\\
$^{1}$Institute for Particle Physics and Astrophysics, ETH Z\"urich, Wolfgang-Pauli-Strasse 27, CH-8093 Z\"urich, Switzerland\\
$^{2}$Swiss Data Science Center, Paul Scherrer Institute, Forschungsstrasse 111, 5232 Villigen, Switzerland\\
}

\date{Accepted XXX. Received YYY; in original form ZZZ}

\pubyear{2025}

\begin{document}
\label{firstpage}
\pagerange{\pageref{firstpage}--\pageref{lastpage}}
\maketitle

\begin{abstract}
Subhalo abundance matching (SHAM) is widely used for connecting galaxies to dark matter haloes.
In SHAM, galaxies and \mbox{(sub-)haloes} are sorted according to their mass (or mass proxy) and matched by their rank order.
In this work, we show that SHAM is the solution of the optimal transport (OT) problem on empirical distributions (samples or catalogues) for any metric transport cost function.
In the limit of large number of samples, it converges to the solution of the OT problem between continuous distributions.
We propose SHAM-OT: a formulation of abundance matching where the halo-galaxy relation is obtained as the optimal transport plan between galaxy and halo mass functions.
By working directly on these (discretized) functions, SHAM-OT eliminates the need for sampling or sorting and is solved using efficient OT algorithms at negligible compute and memory cost.
Scatter in the galaxy-halo relation can be naturally incorporated through regularization of the transport plan.
SHAM-OT can easily be generalized to multiple marginal distributions.
We validate our method using analytical tests with varying cosmology and luminosity function parameters, and on simulated halo catalogues.
The efficiency of SHAM-OT makes it particularly advantageous for Bayesian inference that requires marginalization over stellar mass or luminosity function uncertainties.
\end{abstract}
%
%
%%% Select between one and six entries from the list of approved keywords.
%%% Don't make up new ones.

%%%%%%%%%%%%%%%%%%%%%%%%%%%%%%%%%%%%%%%%%%%%%%%%%%

%%%%%%%%%%%%%%%%% BODY OF PAPER %%%%%%%%%%%%%%%%%%

\section{Introduction}

Dark matter N-body simulations have become an essential tool for understanding the large scale distribution of matter in our universe and its evolution.
However, since these simulations only model dark matter, linking their results to observable galaxies requires the understanding of the galaxy-halo connection (see \cite{wechslerConnectionGalaxiesTheir2018} and references therein).
A commonly used method to populate the haloes from these simulations with galaxies is subhalo abundance matching \citep[SHAM,][]{kravtsovDarkSideHalo2004,conroyModelingLuminositydependentGalaxy2006,valeLinkingHaloMass2004}.
SHAM provides a simple framework for establishing the galaxy-halo connection, while maintaining the necessary complexity through its treatment of substructure.
The SHAM solution is obtained by sorting the galaxy and halo catalogues by their mass (or mass proxy, such as the absolute magnitude) and matching them by their rank order.
Its success in accurately describing this relationship has been demonstrated through multiple applications to observational data \citep{conroyModelingLuminositydependentGalaxy2006,behrooziComprehensiveAnalysisUncertainties2010,guoHowGalaxiesPopulate2010,mosterConstraintsRelationshipStellar2010,wetzelWhatDeterminesSatellite2010,trujillo-gomezGALAXIESLCDMHALO2011,watsonConstrainingSatelliteGalaxy2012,nuzaClusteringGalaxies052013,reddickCONNECTIONGALAXIESDARK2013,simhaTestingSubhaloAbundance2012,chaves-monteroSubhaloAbundanceMatching2016,bernerFastForwardModelling2024}.

To create the SHAM matching function, the galaxy and halo distributions are represented by samples; a large number of samples is required for the matching to be accurate.
This procedure is typically repeated at multiple redshifts.
The generation and sorting of these samples may become computationally expensive for large catalogues, in terms of both time and memory.
While this is manageable for single simulations, it may become a bottleneck for Bayesian inference that marginalizes over uncertainties in the stellar mass or luminosity functions, as the matching has to be re-computed for any new combination of parameters.
Another important element of SHAM is the scatter in the galaxy-halo relation; its inclusion should preserve the halo mass and galaxy luminosity functions. 
For certain models, such as log-normal stellar mass scatter \citep[hereafter \citetalias{behrooziComprehensiveAnalysisUncertainties2010}]{behrooziComprehensiveAnalysisUncertainties2010}, the ``original'' stellar mass function is computed by deconvolving its ``noisy'' counterpart with the scatter model.

In this work, we present SHAM-OT: a novel framework that reformulates the SHAM problem as optimal transport (OT). 
Generally, OT finds a 2D joint probability distribution (or a \emph{transport plan})  such that its marginals are equal to two given 1D distributions, while minimising the cost of transporting one into the other, according to a given cost function \citep[see][for review]{villani2009otoldnew}.
We show that the SHAM procedure is exactly the same as the solution to the optimal transport problem of empirical distributions (represented by samples), under certain assumptions on the transport cost function.
Framing SHAM as OT allows us to operate directly in the large sample limit and use continuous PDF representations of the distributions, instead of samples. 
Standard OT solvers can be used to find the matching from discretised halo and galaxy mass functions (represented as binned histograms, typically with $<\!100$ bins), avoiding the need to process large simulation catalogues (typically $>\!10^{6}$).
In this work we show the benefits of using the SHAM-OT framework: 
(i) computational efficiency, 
(ii) natural inclusion of scatter in the matching relation via OT plan regularization, and 
(iii) straightforward extension to matching more than two distributions using multi-marginal OT.

\section{Optimal Transport}\label{sec:ot_intro}
We briefly introduce the optimal transport (OT) framework, using the Monge-Kantorovich formulation \citep{kantorovich1942translocation, villani2021ot}; see \cite{villani2009otoldnew} for a general overview.
Intuitively, OT is a way to transport mass from one distribution to another while minimizing the cost of transport.

Let's consider \emph{continuous optimal transport}.
For measurable spaces $X$ and $Y$, let 
$a \in \mathcal{P}(X)$ and $b \in \mathcal{P}(Y)$ 
be probability measures on $X$ and $Y$, respectively.
A \emph{transport plan} $\pi$ is a joint probability measure on $X\times Y$ whose marginals on $X$ and $Y$ equal $a$ and $b$, respectively.
The set of all possible transport plans for $a$ and $b$ is 
\begin{equation}
\begin{aligned}
\footnotesize
\Pi(a,b)
\!=\!
\Bigl\{
\,\pi\!&\in\!\mathcal{P}(X \times Y)
: \\
&\int_{A \times Y} \pi(da\,dy)\!=\!a(A),
\
\int_{X \times B} \pi(dx\,db)\!=\!b(B)
\Bigr\},
\end{aligned}
\end{equation}
for all $A\!\subseteq \! X, B\!\subseteq \!Y$ that are also measurable.
Given a cost function $c\!:\!X \times Y\!\to\!\mathbb{R}_{+}$, 
the Kantorovich OT problem finds the \emph{optimal transport plan} $\pi^{*}$ 
that minimizes the total cost
\begin{equation}    
    \pi^* = \arg \min_{\pi \in \Pi(a,b)} \int_{X \times Y} c(x,y) \,\pi(dx\,dy). \label{eq:continuous_ot}
\end{equation}

Now let's consider the \emph{discrete optimal transport}.
For two discrete distributions $\v{p}\!\in\!\mathbb{R}_{+}^n$ and $\v{q}\!\in\!\mathbb{R}_{+}^m$, normalized to $\v{p}\T \mathbf{1}_n\!=\! \v{q}\T \mathbf{1}_m\!=\!1$, where $\v{1}$ is a vector of ones, the transport plan $\v{Q}\! \in \!\Gamma(\v{p},\v{q})$ is a doubly stochastic matrix:
\begin{equation}
\Gamma(\v{p},\v{q}) = \{\v{Q} \in \mathbb{R}_{+}^{n\times m}:  \v{Q} \mathbf{1}_n = \v{p}, \v{Q}\T \mathbf{1}_m = \mathbf{q}\}.
\end{equation}
The cost of the transport is given by $\langle \v{C}, \v{Q} \rangle$, where $\v{C} \in \mathbb{R}_{+}^{n\times m}$ is the given cost matrix and $\langle \cdot, \cdot \rangle$ denotes the Frobenius inner product.
The discrete OT solution is 
\begin{equation}
    \v{Q}^* = \arg \min_{\v{Q} \in \Gamma(\v{p},\v{q})} \langle \v{C}, \v{Q} \rangle.
    \label{eq:discrete_ot}
\end{equation}
This solution may not be unique. 
The problem can be expressed as a linear program and solved using standard solvers.
The transport plan can be penalized by a regularization term, which encourages smoothness of the transport plan and causes the solution to be unique.
The most common penalty is entropic regularization, where a negative entropy term $H(\v{Q})\!=\!\langle \v{Q}, \ln \v{Q} \rangle$ is added:
\begin{equation}
    \v{Q}^*_\epsilon = \arg \min_{\v{Q} \in \Gamma(\v{p},\v{q})} \langle \v{C}, \v{Q} \rangle + \epsilon H(\v{Q}), \label{eq:entropic_ot}
\end{equation}
In the limit of $\epsilon\!\to\!0$, the solution converges to that of the unregularized OT \citep{cominetti1994asymptotic}.
Entropic OT has the additional advantage that it can be solved efficiently using the  Sinkhorn algorithm \citep{sinkhorn1964relationship}.
The Sinkhorn method in Alg.~\ref{alg:sinkhorn} is an iterative algorithm that alternates between row and column normalization.
Its simplicity and parallelizability makes OT applicable to large-scale problems \citep{cuturi2013sinkhorn}.

\begin{algorithm}[H]
\caption{Sinkhorn Algorithm}
\label{alg:sinkhorn}
\begin{algorithmic}[1]
\REQUIRE{Cost matrix $\mathbf{C}$, marginals $\mathbf{p}, \mathbf{q}$, regularization $\epsilon$}
\STATE $\mathbf{K} \leftarrow \exp(-\mathbf{C}/\epsilon)$ \COMMENT{\textit{Compute Gibbs kernel}}
\STATE $\mathbf{u} \leftarrow \mathbf{1}$ \COMMENT{\textit{Initialize scaling vector}}
\REPEAT
    \STATE $\mathbf{v} \leftarrow \mathbf{q}/(\mathbf{K}\mathbf{u})$ \COMMENT{\textit{Scale columns}}
    \STATE $\mathbf{u} \leftarrow \mathbf{p}/(\mathbf{K}^\top\mathbf{v})$ \COMMENT{\textit{Scale rows}}
\UNTIL{convergence}
\RETURN $\text{diag}(\mathbf{u})\mathbf{K}\text{diag}(\mathbf{v})$ \COMMENT{\textit{Compute transport plan}}
\end{algorithmic}
\end{algorithm}

\begin{figure}
    \centering
    \includegraphics[width=1\linewidth]{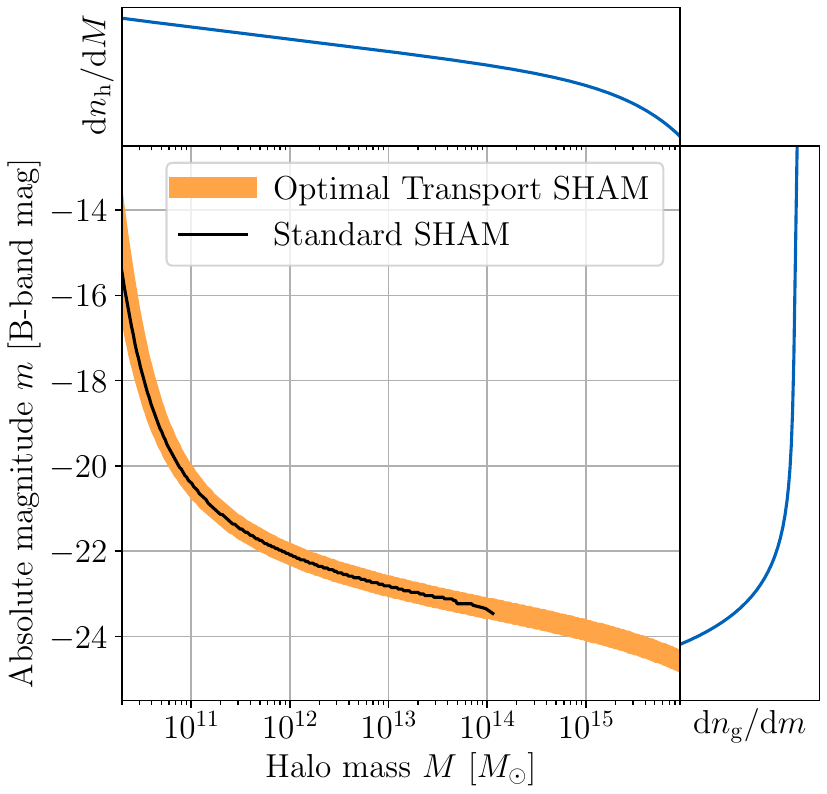}
    \caption{
    Analytical halo mass and luminosity function and the corresponding galaxy-halo connection derived using SHAM-OT and standard SHAM at redshift $z\!=\!0.5$.
    The standard SHAM approach can only be applied up to a halo mass of approximately $10^{14}\solarmass$ due to limited number of massive objects in the sampled catalogue, which here contains $\sim 10^8$ objects.
    In contrast, SHAM-OT operates directly on binned mass and luminosity functions, removing the need for sampling and sorting.
    }
    \label{fig:analytical_match}
\end{figure}

\section{The Subhalo Abundance Matching Problem}\label{sec:sham_intro}

\begin{figure*}
    \centering
    \includegraphics[width=0.9\linewidth]{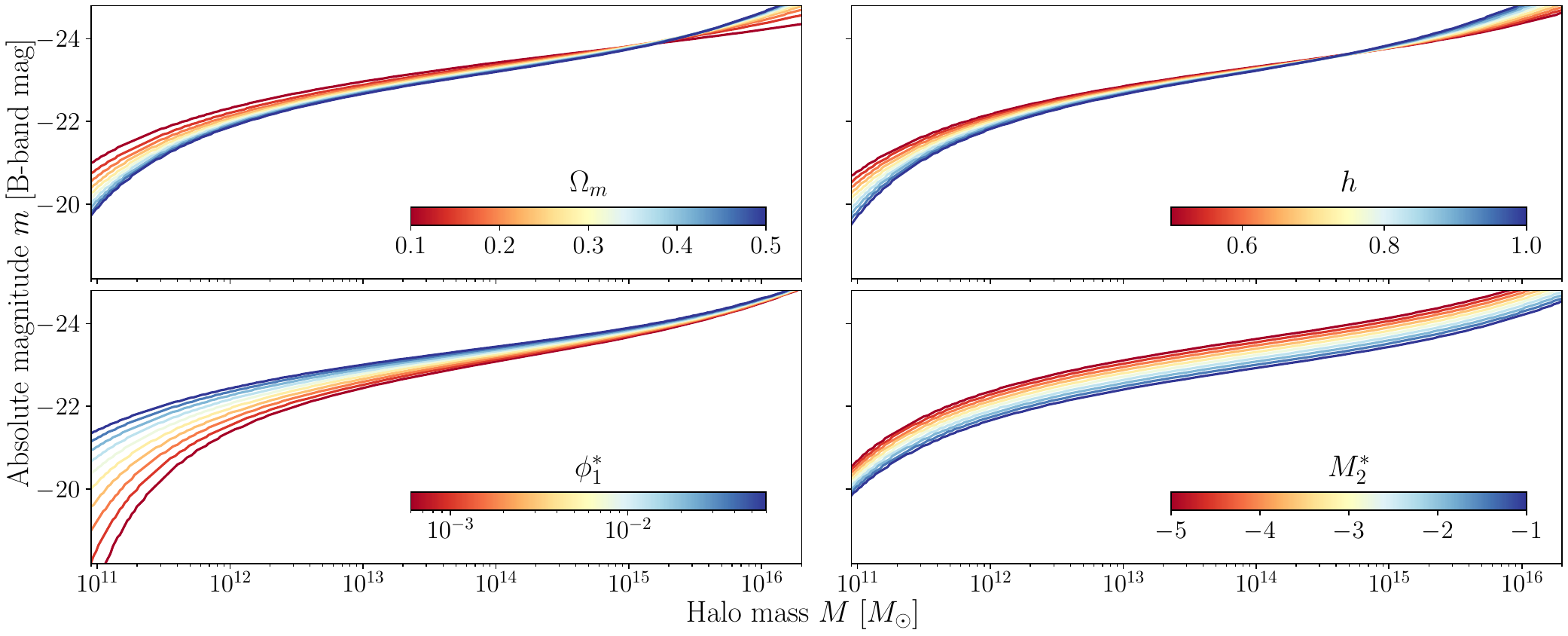}
    \caption{
    Analytical galaxy-halo relation and their dependence on different parameters at redshift $z\!=\!0.5$.
    We show the impact of the matter density $\Omega_m$ (top left), the Hubble parameter $h$ (top right), the amplitude of the luminosity function at redshift $z\!=\!0$ given by $\phi_1^*$ (bottom left) and $M_2^*$ which is responsible for the redshift evolution of $M^*$, therefore changing the redshift evolution of the knee of the luminosity function.
    }
    \label{fig:vary}
\end{figure*}

Abundance matching techniques find the relationship between galaxies and dark matter haloes.
In their simplest form, these methods assume a monotonic relationship between the halo mass and a given galaxy property, such as stellar mass or absolute magnitude.
The galaxy-halo connection is then obtained by ranking both galaxies and haloes by these properties, and matching the most massive (or brightest) galaxies with the most massive haloes.
Subhalo abundance matching (SHAM) extends this method to account for subhaloes and produces a relation that agrees better with observations \citep{kravtsovDarkSideHalo2004,conroyModelingLuminositydependentGalaxy2006,valeLinkingHaloMass2004}.
Additionally, more realistic models incorporate scatter in the galaxy-halo relation \citep{behrooziComprehensiveAnalysisUncertainties2010, trujillo-gomezGALAXIESLCDMHALO2011, reddickCONNECTIONGALAXIESDARK2013, zentnerGalaxyAssemblyBias2014}. 
For example, \citetalias{behrooziComprehensiveAnalysisUncertainties2010} demonstrates how log-normal scatter of stellar mass at a fixed halo mass can be added, while preserving the marginal distribution of the stellar mass function.
Furthermore, differentiating between passive red galaxies and star-forming blue galaxies enhances modelling accuracy, as these populations exhibit distinct preferences for occupation in haloes versus subhaloes \citep{dragomirDoesGalaxyHalo2018, rodriguez-pueblaStelLARTOHALOMASSRELATION2015,Masaki_Lin_Yoshida_2013, yamamotoTestingSubhaloAbundance2015,bernerFastForwardModelling2024}.

The core computational challenge is the same for all the SHAM models listed above: sorting two catalogues to establish the galaxy-halo relation for each redshift interval.
The accuracy of matching relies on using very large catalogues, possibly with billions of objects, which becomes computationally intensive in terms of computing time and memory footprint, especially if re-computed frequently.

\section{SHAM as Optimal Transport}\label{sec:sham-ot}

We show that the SHAM procedure can be expressed as optimal transport, and provide a sketch of the proof using basic results from OT and statistics theory.
Let $\{x_i\}_N{\sim}a$ and $\{y_i\}_N{\sim}b$ be i.i.d.\ samples from two distributions.
In the case of SHAM, these correspond to the halo and the galaxy catalogues, respectively.
An empirical distribution $\pi_N^\sigma \in \mathcal{P}(X \times Y)$ created by pairing the samples is
\begin{equation}
    \pi_N^\sigma = \frac{1}{N}\sum_{i=1}^N \delta(x_i, y_{\sigma(i)}),
\end{equation}
where $\sigma{\in}I!$ is a permutation sequence of the indices $I{=}\{1, \ldots, N\}$, and $\delta$ is a Dirac measure.
Let us define the metric cost function $C(x,y)$ with the metric properties: $C(0,0)\!=\!0, C(x,y)\!\geq\!0, C(x,y)\!=\!C(y,x), C(x,y)\!\leq\!C(x,z)\! + \!C(z,y) \ \forall \ x\!\in\!X, y\!\in\!Y, z\!\in\!Z.$
The goal is to find a permutation $\sigma$ that minimizes this cost, similarly to Eqn.~\ref{eq:continuous_ot}
\begin{equation}
    \pi_N^{*} = \arg \min_{\sigma \in I!} \int_{X \times Y} C(x,y) \pi_N^\sigma(dx\,dy).
\end{equation}
Let $x^*_k, y^*_k$ be the $k$-th order statistics, such that the samples are sorted in ascending order
$
    x_1^*\!\leq\!x_2^*\!\leq\!\cdots\!\leq\!x_N^*, \ y_1^*\!\leq\!y_2^*\!\leq\!\cdots\!\leq\!y_N^*,
$
For the metric cost $C(\cdot, \cdot)$, the optimal transport plan $\pi_N^*$ is given by matching the order statistics \citep[Lemma 4.2]{bobkov2019kantorovich}:
\begin{equation}
    \pi_N^* = \frac{1}{N}\sum_{k=1}^N \delta(x_k^*, y_k^*) \in \mathcal{P}(X \times Y),
\end{equation}
Intuitively, for the sorted matched samples, the cost cannot be lowered by exchanging any two matches.
Theorem 4.3 in \citet{bobkov2019kantorovich} establishes the approximation bounds for transport cost between the empirical distribution and the true distribution.
By the Fundamental Theorem of Statistics (Glivenko–Cantelli), the empirical distribution converges to the true distribution $\pi^*_N \to \pi^*$ when $N\to\infty$ almost surely.
Therefore, in the limit of $N\to\infty$, the solution to the SHAM procedure is equal to the solution of the continuous optimal transport in Eqn.~\ref{eq:continuous_ot} with any metric cost function. 

\begin{figure*}

    \begin{minipage}[t]{0.48\linewidth}
        \centering
        \includegraphics[width=1\linewidth]{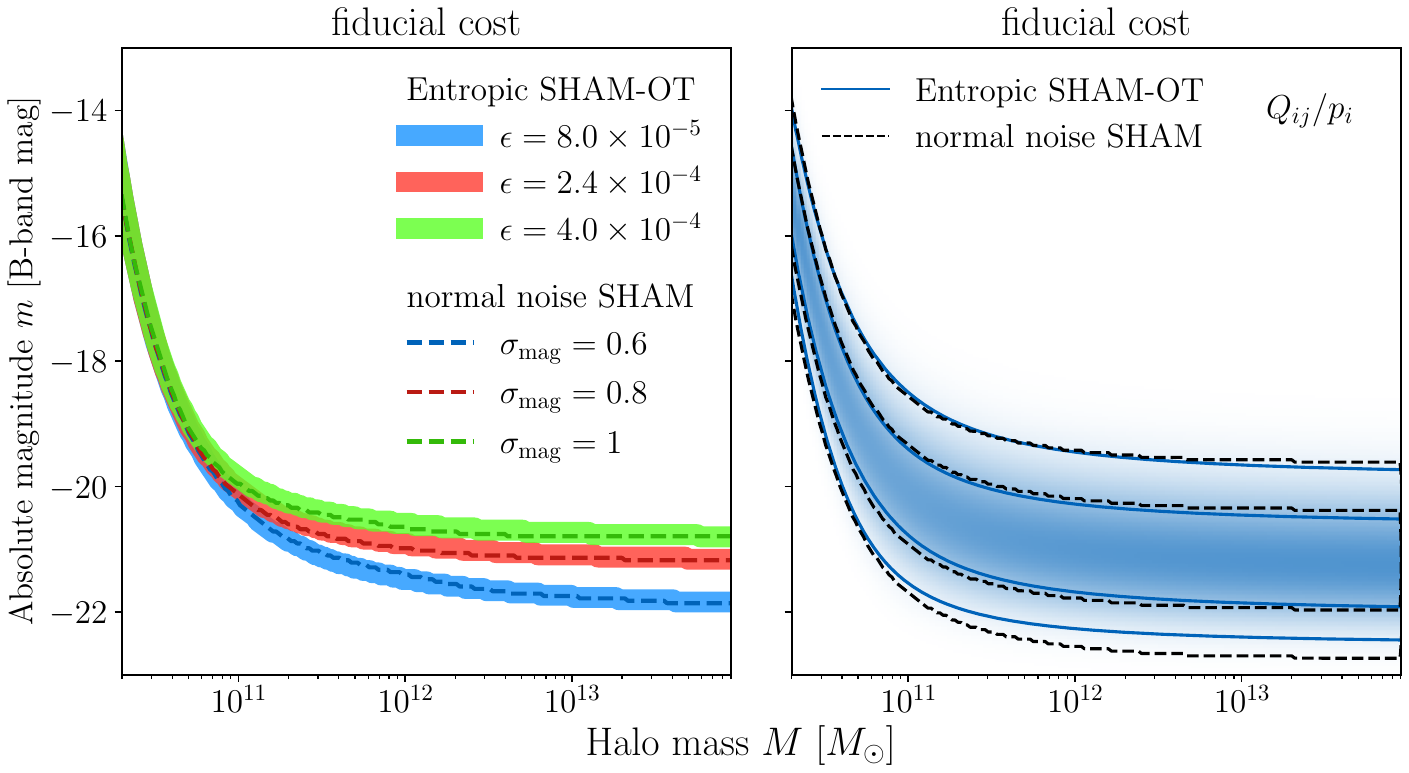}
        \caption{
        Comparison of SHAM-OT with entropic regularization and standard SHAM.
        Left: Mean magnitude vs.\ halo mass for different $\epsilon$ and $\sigma_\mathrm{mag}$.
        Right: Transport plan $\v{Q}$ normalized by halo marginal $\v{p}$, shown for the middle case on the left.
        SHAM-OT (blue) agrees well with standard SHAM (black).
        }
        \label{fig:reg_sham}
    \end{minipage}
    \hfill
    \begin{minipage}[t]{0.48\linewidth}
        \centering
        \includegraphics[width=1\linewidth]{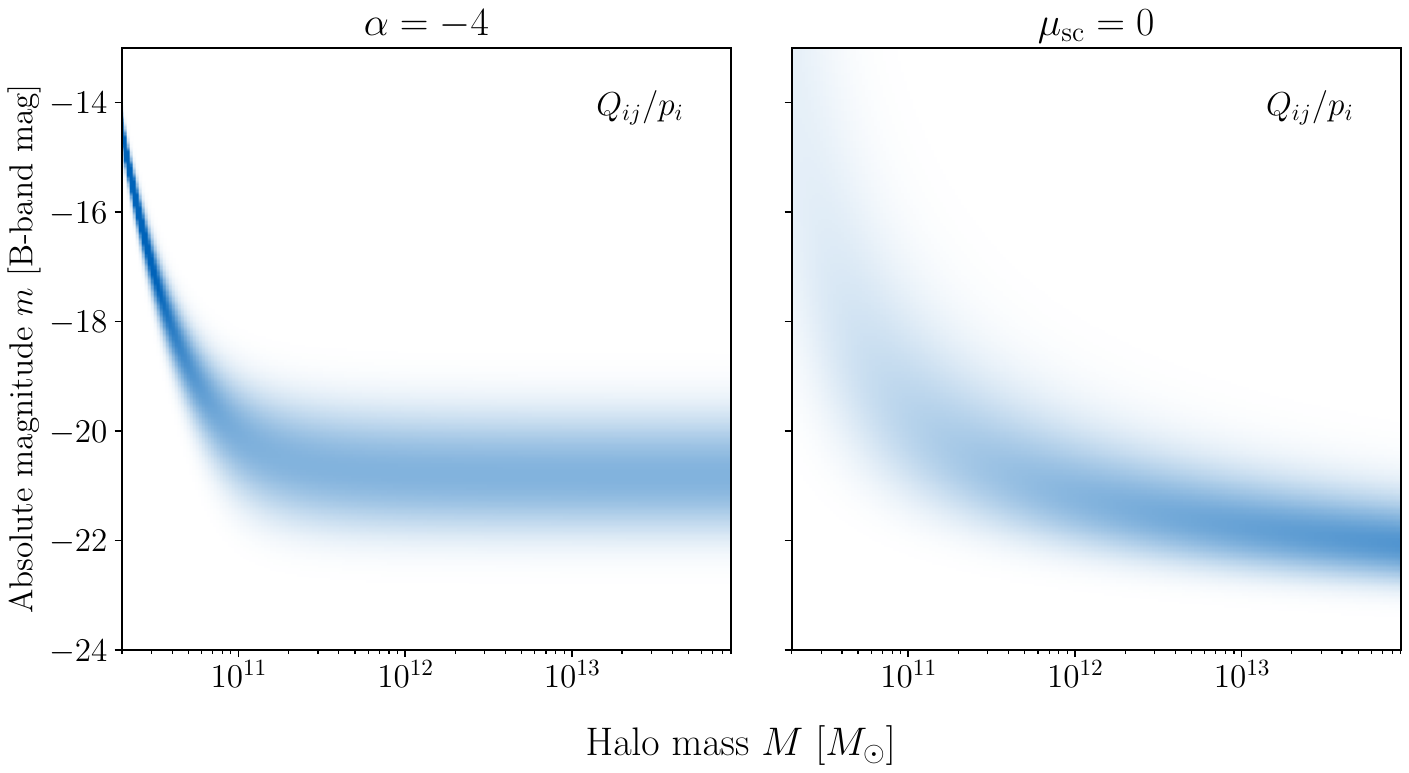}
        \caption{
        Impact of cost function choices on the transport plan.
        The regularization parameter $\epsilon$ is tuned to reproduce the same scatter at $M = 10^{12}M_\odot$ as in Figure \ref{fig:reg_sham}.
        Left: Using exponent $\alpha = -4$ concentrates the plan at low halo masses.
        Right: Setting $\mu_\mathrm{sc} = 0$ increases scatter at low masses.
        }
        \label{fig:different_regularization}
    \end{minipage}
\end{figure*}

\section{Matching halo mass and galaxy luminosity with Optimal Transport}
\label{sec:analytical_match}

We demonstrate how optimal transport can be used to derive the galaxy-halo relation from analytical distributions.
The halo mass function gives the number of haloes $n_\mathrm{h}$ per mass $M$
\begin{equation}
\small
\frac{\mathrm{d}n_\mathrm{h}}{\mathrm{d}M} = f(\sigma) \frac{\bar{\rho}}{M} \frac{\mathrm{d} \ln \sigma^{-1}}{\mathrm{d}M},
\label{eq:halo_mass_func}
\end{equation}
where $\sigma$ is the variance of a spherical top hat containing the mass $M$, $\bar \rho$ is the mean matter density and $f(\sigma)$ is the halo multiplicity function.
We use the multiplicity function by \cite{watsonHaloMassFunction2013} which is given by
\begin{equation}
\small
f(\sigma) = A \left[ \left( \frac{\beta}{\sigma} \right)^{\alpha} + 1 \right] \exp \left( - \frac{\gamma}{\sigma^2} \right),
\end{equation}
with the parameters $A\!=\!0.282$, $\alpha\!=\!2.163$, $\beta\!=\!1.406$, and $\gamma\!=\!1.210$.
The subhalo mass function, for a subhalo of mass $M_\mathrm{s}$, is given by
\begin{equation}
\frac{\mathrm{d}n}{\mathrm{d} \ln (M_\mathrm{s}/M)} = \frac{f_0}{\beta \Gamma(1 - \gamma)} \left( \frac{M_\mathrm{s}}{\beta M} \right)^{-\gamma} \exp \left( - \frac{M_\mathrm{s}}{\beta M} \right),
\label{eq:subhalo_mass_func}
\end{equation}
where $M$ is the central halo mass, and the parameters are $\beta\!= \!0.3$, $\gamma\!=\!0.9$, and $f_0\!=\!0.1$, following \cite{moGalaxyFormationEvolution2010}.
The subhalo mass function is computed as a weighted average of the subhalo mass functions over the entire range of central halo masses considered, using the halo mass function as the weight at each mass.
This yields the full subhalo mass function, which is then combined with the halo mass function to derive the dark matter host mass function.

Popular choices for the galaxy mass proxy include the stellar mass and the absolute magnitude (see \cite{wechslerConnectionGalaxiesTheir2018} for an overview).
In this work, we use the absolute magnitude in the B-band in the parametrization from \cite{fischbacherGalSBIPhenomenologicalGalaxy2024} (hereafter \citetalias{fischbacherGalSBIPhenomenologicalGalaxy2024}).
The number of galaxies $n_\mathrm{g}$ per absolute magnitude $m$ is given by
\begin{equation}
    \label{eq:lumfunc}
    \small
    \frac{\mathrm{d}n_\mathrm{g}}{\mathrm{d}m} = \frac{2}{5} \ln(10) \phi^*(z) 10^{\frac{2}{5} (M^*(z) - M)(\alpha+1)} \exp\left(-10^{\frac{2}{5}\left(M^*\left(z\right)-M\right)}\right),
\end{equation}
where $\alpha$ is a free parameter and $\phi^*(z)$ and $M^*(z)$ are functions of redshift parametrized by
\begin{equation*}\label{eq:trunc_logexp}
\small
    \phi^*(z) = \phi^*_1 \exp(\phi^*_2 z), \ \
    M^*(z) = 
    \begin{cases} 
        M^*_1 + M^*_2 \log(1+z) & \text{if } z < z_0 \\
        M^*_1 + M^*_2 \log(1+z_0) & \text{if } z \geq z_0
    \end{cases},
\end{equation*}
where $M^*_1, M_2^*, \phi^*_1, \phi^*_2$ and $z_0$ are free parameters.
The free parameters are fixed to the first index of the blue population of the publicly available model by \citetalias{fischbacherGalSBIPhenomenologicalGalaxy2024}.

The transport of distributions requires them to have the same total mass.
We ensure this by defining a lower limit on the host mass:
we choose $M^\mathrm{min}\!=\!10^{10}\!M_{\odot}$, and then compute the necessary faintest absolute magnitude such that the number of haloes and galaxies per unit volume is equal.
For numerical reasons, we also define an upper limit on both the halo mass ($M^\mathrm{max}{=}10^{19} M_{\odot}$) and the brightest absolute magnitude ($m^\mathrm{min}$=-27).
These numbers are chosen such that the number of haloes or galaxies above those limits is so small that it does not impact the matching results from SHAM.
We create a linearly-spaced vector of  negative magnitudes $\vg{\nu}\!\in\!\mathbb{R}^{n}\!:\!\nu_i\!=\!-m_i$ and logarithmically-spaced halo mass vector $\vg{\mu}\!\in\!\mathbb{R}^{n}\!:\! \mu_i\!=\!\log_{10} M_{i}$, both with $n\!=\!500$ within the limits described above.
%, resulting in sampling spacing of $\Delta \tilde{\nu}\!=\! 0.036$ and $\Delta \tilde{\mu}\!=\! 0.018$.
We generate the discretized halo mass function $\v{p}\!\in\!\mathbb{R}_+^{n}$ using equations~\ref{eq:halo_mass_func} and \ref{eq:subhalo_mass_func}, and the galaxy absolute magnitude function $\v{q} \in \mathbb{R}_+^{n}$ using Eqn.~\ref{eq:lumfunc}.
We show $\v{p}$ and $\v{q}$ in side panels in Fig.~\ref{fig:analytical_match}.
Since the choice of cost metric does not impact the unregularized optimal transport plan, we use $C_{ij}\!=\!(\nu_i\!-\!\mu_j)^2$ and obtain the optimal transport plan by solving the problem in Eqn.~\ref{eq:discrete_ot} using linear programming.
In the standard SHAM approach, we sample catalogues in an interval $z\!\in\![0.495,0.505]$, which is typical for practical abundance matching, resulting in $\sim \! 10^8$ objects.
The galaxy-halo connection can only be computed up to a mass limit of about $10^{14} \solarmass$, as the rapid decline of the mass functions prevents any higher-mass haloes in the simulated catalogue.

The resulting optimal transport plan $\v{Q}$ is presented in Fig.~\ref{fig:analytical_match} and compared with the standard SHAM approach: they align almost perfectly.
Residual differences arise because standard SHAM uses finite simulation catalogues, where massive halos are few and subject to sampling noise -- especially at the high-mass end.
In contrast, SHAM-OT works with binned distributions, allowing smooth extension to higher halo masses without discreteness effects.
Furthermore, we find a computational speed improvement of $O(100)$ for SHAM-OT compared to standard SHAM.
SHAM-OT's computational efficiency allows us to easily study the impact of changes in the luminosity or halo mass function.
In Fig.~\ref{fig:vary}, we show how different values of the cosmological and luminosity function parameters impact the galaxy-halo connection.

\begin{figure*}
    \centering
    \includegraphics[width=0.8\linewidth]{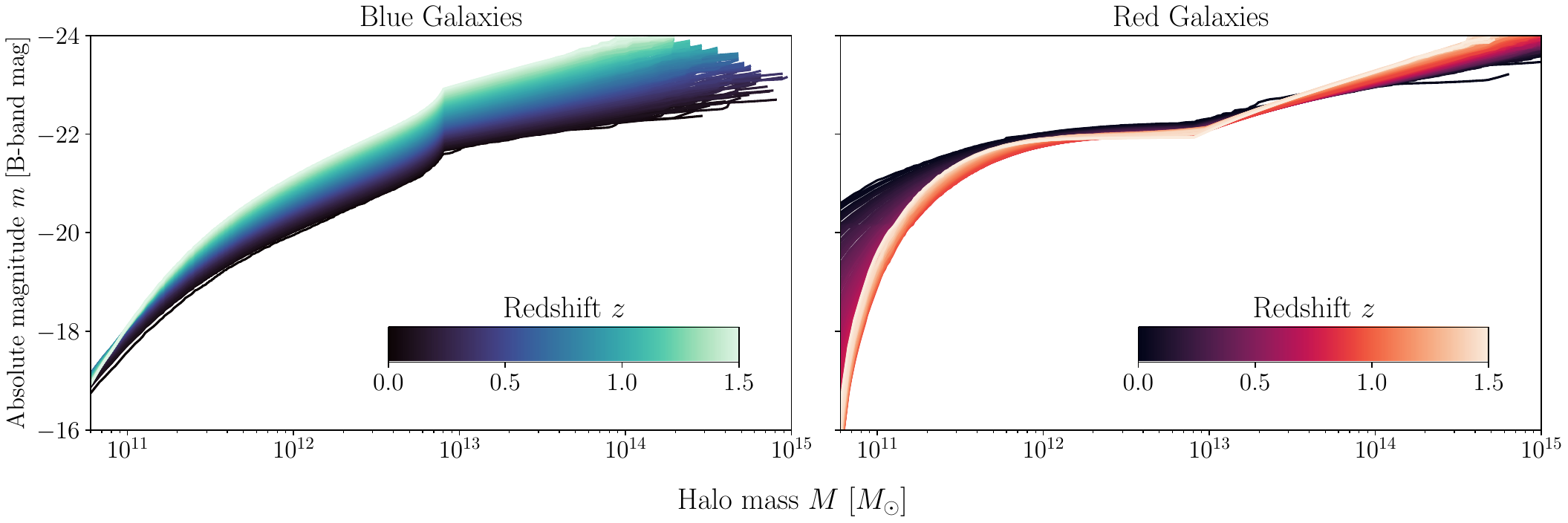}
    \caption{Galaxy-halo relation for red and blue galaxies in a redshift range from $z\!=\!0$ to $z\!=\!1.5$. The relatively abrupt change in the relation at $8\!\times\! 10^{12}M_{\odot}$ is due to the sharp mass limit used for the red/blue assignment.}
    \label{fig:sims}
\end{figure*}

\section{Entropic Regularization as a Scatter Model}\label{sec:entropic-cosmic-variance}

In SHAM, the scatter noise is often modelled by a conditional probability distribution on the stellar mass given a fixed halo mass \citepalias{behrooziComprehensiveAnalysisUncertainties2010}.
A common choice is a log-normal distribution, the scale of which may vary with halo mass.
This noise is added after the matching procedure.
To ensure the preservation of the stellar mass distribution after adding the noise, the ``direct'' stellar mass function is used for matching, which is created by deconvolving the ``true'' stellar mass function with the desired noise model \citepalias{behrooziComprehensiveAnalysisUncertainties2010}.
This deconvolution step may cause difficulties at high-mass end with low number of objects, as well as for more advanced noise models.

In the OT framework, the marginals of the transport plan are guaranteed to match the input distributions.
The effect of scatter can be modelled naturally by the use of regularization of the transport plan: 
increasing the regularization parameter $\epsilon$ will result in a smoother plan.
For the regularized OT, the choice of cost metric directly affects the structure of the resulting transport plan.
Various noise model behaviours, including the dependence on halo mass or galaxy luminosity, can therefore be designed through different choices of the cost function, as long as it satisfies the metric requirements.

We demonstrate the capacity of the entropic regularization to model various scattering properties, including ones that mimic the log-normal approach.
We use the same analytical prescription with the same cuts for the halo mass and luminosity function as described in Sec.~\ref{sec:analytical_match},
and make sure that potential edge effects at the lower mass limit have negligible impact on the transport plan.
We compute the standard SHAM solution with Gaussian scatter in magnitude, using the method from \citetalias{behrooziComprehensiveAnalysisUncertainties2010}
for three values of scatter $\sigma_{\mathrm{mag}}$.
To properly emulate this noise behaviour with SHAM-OT, we adapt the basic cost function from the previous section by two parameters $\mu_\mathrm{sc}$ and $\alpha$ giving $C_{ij}\!=\!(\nu_i\!-\!\mu_j)^2/(\mu_j\!-\!\mu_\mathrm{sc})^\alpha$.
We use $\mu_\mathrm{sc}\!=\!9.94$ and $\alpha\!=\!-1.7$ for our fiducial setup.
The OT solution is obtained with the Sinkhorn algorithm (Alg.~\ref{alg:sinkhorn}).
Fig. \ref{fig:reg_sham} shows the comparison between the two approaches for different noise levels and corresponding $\epsilon$ for our fiducial case.
SHAM-OT matches the log-normal model well, both in terms of the mean of the distribution (left panel) and its shape (right panel) and is able to reproduce the behaviour of the standard SHAM.

Beyond emulating the log-normal scatter, regularized OT offers a flexible framework to model different scatter behaviours.
Fig.~\ref{fig:different_regularization} illustrates this flexibility by comparing the conditional transport plan for two different choices of cost functions.
Further exploration of cost functions, along with the development of physically motivated choices to reproduce existing scatter behaviour with high precision, is left for future work.

\section{SHAM-OT with haloes from simulations}\label{sec:pinocchio-sham}

In practice, the halo mass function is often not analytically given, but can instead be computed from empirical halo distributions based on simulations.
We use the approximate code \texttt{PINOCCHIO} \citep{monacoPINOCCHIOPinpointingOrbitcrossing2002,taffoniPINOCCHIOHierarchicalBuildup2002,monacoPredictingNumberSpatial2002,monacoAccurateToolFast2013,munariImprovingFastGeneration2017,rizzoSimulatingCosmologies$L$CDM2017} to simulate the past light cone of dark matter haloes.
\texttt{PINOCCHIO} uses Lagrangian perturbation theory as an approximation which makes it faster than full N-body simulations.
We assign subhaloes to the halo catalogue based on the merger tree using the method described in \cite{bernerRapidSimulationsHalo2022} and label the haloes and subhaloes as hosts for red or blue galaxies based on the model from \cite{bernerFastForwardModelling2024}.
This model is based on two key parameters: the halo mass limit $ M_{\text{limit}} \!=\! 8 \!\times\! 10^{12} \, h^{-1} M_{\odot} $ and the satellite quenching time $ t_{\text{quench}} \!=\! 2.0 $ Gyr.
Central haloes with masses above $ M_{\text{limit}} $ host red galaxies, while lower-mass haloes host blue galaxies.
Galaxies in subhaloes remain blue until they have spent more than $ t_{\text{quench}} $ within their host, after which they transition to red.
As a result, we obtain catalogues of haloes and subhaloes with their corresponding mass, position in 3D space, and population label (i.e.\ red or blue).
Note that the methods above can easily be replaced by any other approach that provides a halo catalogue containing the same parameters.
We split the host catalogue into redshift bins with width $\Delta z\!=\!0.01$. For each bin, we obtain a host mass function as a histogram, including the upper (heaviest host in the redshift bin) and lower (mass resolution of the simulation) mass limits.
We then utilize the luminosity function as defined in \citetalias{fischbacherGalSBIPhenomenologicalGalaxy2024} with the parameters for the red and blue population from the first index of the posterior distribution.
Similar to the procedure described in Sec.~\ref{sec:analytical_match}, we determine the faintest absolute magnitude required such that the number of hosts equals the expected number of galaxies in this redshift bin.
Using the constructed distributions for the host mass and galaxy absolute magnitudes, we compute the OT solution for each redshift bin, for both red and blue galaxies.
The resulting galaxy-halo relation at different redshifts, for both red and blue galaxies, is shown in Fig.~\ref{fig:sims}.
Introducing scatter to this relation would be straightforward, however a careful choice of the cost matrix would then be necessary, see also Section \ref{sec:entropic-cosmic-variance}.

\section{Matching more than two functions with multi-marginal optimal transport}\label{sec:multimarginal}

\begin{figure}
    \centering
    \includegraphics[width=0.9\linewidth]{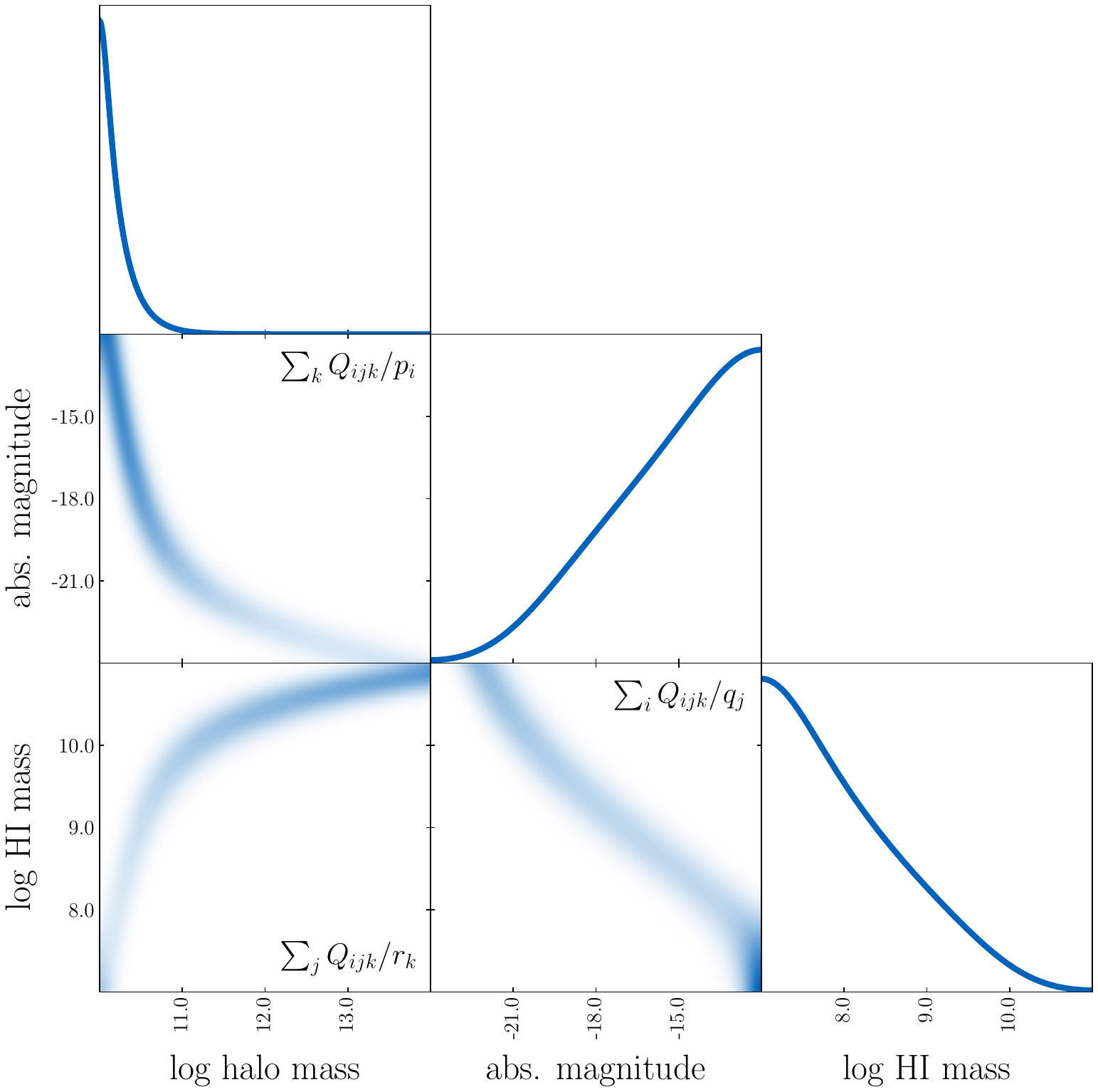}
    \caption{
    Conditional distributions from the multi-marginal optimal transport plan $Q$ with halo mass $p$, absolute magnitude $q$, and HI mass $r$, using $\epsilon{=}0.001$. Density panels show conditional transport plans $\sum_k Q_{ijk}/p_i$, $\sum_i Q_{ijk}/q_j$, $\sum_j Q_{ijk}/r_k$.
    The 1D marginals are preserved by the transport plan by construction.
    }
    \label{fig:multi-marginal}
\end{figure}

Multi-marginal optimal transport is a generalization of optimal transport that applies to any number of marginal distributions \citep[see e.g.][]{benamou2015iterative,pass2015multimargot}.
We will consider three distributions in the discrete setting, but the method can be extended analogously to higher dimensions.
Given distributions $\v{p} \!\in\!\mathbb{R}_{+}^{n}, \v{q} \!\in\!\mathbb{R}_{+}^{m}, \v{r} \!\in\!\mathbb{R}_{+}^{k}$ such that  $\v{p}\T\v{1}\!=\v{q}\T\v{1}\!=\v{r}\T\v{1}\!=\!1$ and a cost tensor $\v{C} \!\in\!\mathbb{R}_{+}^{n \times m \times k}$, entropy-regularized multi-marginal OT finds the transport plan $\v{Q} \!\in\!\mathbb{R}_{+}^{n \times m \times k}$ that minimizes the cost
\begin{equation}
     \min_{\v{Q} \in \Gamma(\v{p}, \v{q}, \v{r})} \langle \v{C}, \v{Q} \rangle + \epsilon H(\v{Q}),
\end{equation}
where the set of allowed transport plans is
\small
\begin{equation*}
    \Gamma(\v{p}, \v{q}, \v{r})\!=\! \Bigl\{ \v{Q}\!\in\!\mathbb{R}_{+}^{n \times m \times k} : \sum_{jk} Q_{ijk}\!=\!p_i, \sum_{ik} Q_{ijk}\!=\!q_j, \sum_{ij} Q_{ijk}\!=\!r_k \Bigl\}
\end{equation*}
\normalsize
The Sinkhorn algorithm, extended to this setting \citep{benamou2015iterative}, is shown in Alg.~\ref{alg:multimarginal}.
We demonstrate the multi-marginal matching, including scatter, between three distributions: halo mass, galaxy luminosity, and HI mass.
The HI mass function is given by 
\begin{equation}
    \label{eq:h1mass}
    \frac{\mathrm{d}n_\mathrm{HI}}{\mathrm{d}M} = \ln(10)  \phi_* \left(\frac{M_\mathrm{HI}}{M_*}\right)^{\alpha+1} \exp\left( -\frac{M_\mathrm{HI}}{M_*} \right)
\end{equation}
with $M_*\!=\!10^{9.94}, \phi_*\!=\!4.5\times 10^{-3}\,\mathrm{Mpc}^{-3} \mathrm{dex}^{-1}$ and $\alpha\!=\!-1.25$ following \cite{jones2018alfalfa}.
Similar to before, we make sure that the lower mass limit of HI mass is chosen such that the three matched distributions are normalized consistently.
The multi-marginal matching plan is shown in Fig.~\ref{fig:multi-marginal}.
The 1D marginals are preserved by construction, eliminating the need for complex deconvolution methods required in standard SHAM.

\begin{algorithm}
    \begin{algorithmic}
        \REQUIRE{Cost tensor $\mathbf{C}$, marginals: $\mathbf{p}$, $\mathbf{q}$, $\mathbf{r}$, regularization: $\epsilon$}
        \STATE $\mathbf{K} \leftarrow \exp(-\mathbf{C}/\epsilon)$ \COMMENT{\textit{Compute Gibbs kernel}}
        \STATE $\mathbf{u} \leftarrow \mathbf{1}, \mathbf{v} \leftarrow \mathbf{1}, \mathbf{w} \leftarrow \mathbf{1}$ \COMMENT{\textit{Initialize scaling vectors}}
        \REPEAT \STATE \vspace{-2em} \begin{talign*}
                u_i &\leftarrow p_i / \sum_{jk} K_{ijk} v_j w_k \\[-2pt]
                v_j &\leftarrow q_j / \sum_{ik} K_{ijk} u_i w_k \\[-2pt]
                w_k &\leftarrow r_k / \sum_{ij} K_{ijk} u_i v_j 
            \end{talign*}
        \vspace{-1.7em}
        \UNTIL{convergence}
        \RETURN $Q_{ijk} = K_{ijk} u_i v_j w_k$ \COMMENT{\textit{Compute transport plan}}
    \end{algorithmic}
    \caption{Sinkhorn algorithm for 3D multi-marginal OT}
    \label{alg:multimarginal}
\end{algorithm}

\section{Conclusions and discussion}\label{sec:conclusions}

We show that the SHAM sort-matching algorithm converges to the solution of the continuous optimal transport problem in the limit of large number of samples. 
By employing the discrete OT formulation, where the distributions are represented as probability mass functions instead of samples, the matching relation (transport plan) can be found with standard OT solvers at negligible compute and memory cost. 
Therefore, we only need to solve a problem whose complexity depends on the number of bins in the probability mass functions -- typically on the order of $O(100)$ -- rather than on the number of samples, which can be as large as $O(>\!10^6)$.

We present the SHAM-OT framework, which straightforwardly extends beyond the standard SHAM hard-matching of two distributions: 
(i) it can naturally include scatter models via regularization, and 
(ii) it can match more than two distributions in the multi-marginal variant.
We show the agreement of OT and SHAM in realistic demonstration examples of matching the halo mass function and the galaxy luminosity function, from both analytical prescription for the halo mass function and from a simulated halo catalogue.
We demonstrate how regularized OT can be used to create scatter on the mass-luminosity relation, while satisfying the marginal constraints by construction.
Finally, we show that SHAM-OT can be used to match more than two functions, through an example of matching halo mass, galaxy luminosity, and HI mass functions.
The SHAM-OT provides instantaneous solutions to the abundance matching problems, which presents multiple benefits for generating simulations, especially if the matching is recomputed frequently for different parameters of luminosity or HI mass functions.

The SHAM-OT framework can be further studied in several directions. 
The properties of scatter can further be modelled by modifying the metric transport cost.
It can potentially be closely linked with existing models of log-normal scatter, as well as be used to straightforwardly explore new models. 
The regularization strength $\epsilon$ controlling the amount of scatter can be easily included as part of the model and marginalized over a prior in Bayesian inference.

While the entropic penalty is the most commonly used in the OT community, other regularisers can also implemented to create new scatter models.
Quadratic OT \citep{benamou2000computational, nutz2024quadratically}, which penalizes the transport plan by its squared Frobenius norm $\| Q \|_F^2$, is particularly interesting for this as it has deep connections with 
fluid dynamics \citep{benamou2000computational}, 
Lagrangian reconstruction methods for large-scale structure \citep{brenier2003reconstruction},
and thermal diffusion \citep{jordan1998fokkerplanck}.

\section*{Acknowledgements}

We acknowledge the use of the following software packages: \texttt{numpy} \citep{vanderwaltNumPyArrayStructure2011}, \texttt{galsbi} \citep{fischbachergalsbiPythonPackage2024}, \texttt{POT} \citep{flamaryPOTPythonOptimal2021}, \texttt{PyCosmo} \citep{refregierPyCosmoIntegratedCosmological2018,tarsitanoPredictingCosmologicalObservables2021,moserSymbolicImplementationExtensions2022,hitzFastSimulationCosmological2024}, \texttt{scipy} \citep{virtanenSciPy10FundamentalAlgorithms2020}.
Plots were created using \texttt{matplotlib} \citep{hunterMatplotlib2DGraphics2007} and \texttt{trianglechain} \citep{fischbacherRedshiftRequirementsCosmic2023,kacprzakDeepLSSBreakingParameter2022}.
This project was supported in part by grant 200021\_192243 from the Swiss National Science Foundation.
The authors acknowledge support from SERI and the SKACH consortium.

\section*{Software and Data Availability}

The code to reproduce this analysis is available under this link: \url{https://cosmo-gitlab.phys.ethz.ch/cosmo_public/sham-ot/}. The simulation data can be made available upon request.

%%%%%%%%%%%%%%%%%%%% REFERENCES %%%%%%%%%%%%%%%%%%

\bibliographystyle{mnras}
\bibliography{bibliography}

\begin{thebibliography}{}
\makeatletter
\relax
\def\mn@urlcharsother{\let\do\@makeother \do\$\do\&\do\#\do\^\do\_\do\%\do\~}
\def\mn@doi{\begingroup\mn@urlcharsother \@ifnextchar [ {\mn@doi@} {\mn@doi@[]}}
\def\mn@doi@[#1]#2{\def\@tempa{#1}\ifx\@tempa\@empty \href {http://dx.doi.org/#2} {doi:#2}\else \href {http://dx.doi.org/#2} {#1}\fi \endgroup}
\def\mn@eprint#1#2{\mn@eprint@#1:#2::\@nil}
\def\mn@eprint@arXiv#1{\href {http://arxiv.org/abs/#1} {{\tt arXiv:#1}}}
\def\mn@eprint@dblp#1{\href {http://dblp.uni-trier.de/rec/bibtex/#1.xml} {dblp:#1}}
\def\mn@eprint@#1:#2:#3:#4\@nil{\def\@tempa {#1}\def\@tempb {#2}\def\@tempc {#3}\ifx \@tempc \@empty \let \@tempc \@tempb \let \@tempb \@tempa \fi \ifx \@tempb \@empty \def\@tempb {arXiv}\fi \@ifundefined {mn@eprint@\@tempb}{\@tempb:\@tempc}{\expandafter \expandafter \csname mn@eprint@\@tempb\endcsname \expandafter{\@tempc}}}

\bibitem[\protect\citeauthoryear{Behroozi, Conroy  \& Wechsler}{Behroozi et~al.}{2010}]{behrooziComprehensiveAnalysisUncertainties2010}
Behroozi P.~S.,  Conroy C.,   Wechsler R.~H.,  2010, \mn@doi [The Astrophysical Journal] {10.1088/0004-637X/717/1/379}, 717, 379

\bibitem[\protect\citeauthoryear{Benamou \& Brenier}{Benamou \& Brenier}{2000}]{benamou2000computational}
Benamou J.-D.,  Brenier Y.,  2000, Numerische Mathematik, 84, 375

\bibitem[\protect\citeauthoryear{Benamou, Carlier, Cuturi, Nenna  \& Peyr{\'e}}{Benamou et~al.}{2015}]{benamou2015iterative}
Benamou J.-D.,  Carlier G.,  Cuturi M.,  Nenna L.,   Peyr{\'e} G.,  2015, SIAM Journal on Scientific Computing, 37, A1111

\bibitem[\protect\citeauthoryear{Berner, Refregier, Sgier, Kacprzak, Tortorelli  \& Monaco}{Berner et~al.}{2022}]{bernerRapidSimulationsHalo2022}
Berner P.,  Refregier A.,  Sgier R.,  Kacprzak T.,  Tortorelli L.,   Monaco P.,  2022, \mn@doi [Journal of Cosmology and Astroparticle Physics] {10.1088/1475-7516/2022/11/002}, 2022, 002

\bibitem[\protect\citeauthoryear{Berner, Refregier, Moser, Tortorelli, Machado Poletti~Valle  \& Kacprzak}{Berner et~al.}{2024}]{bernerFastForwardModelling2024}
Berner P.,  Refregier A.,  Moser B.,  Tortorelli L.,  Machado Poletti~Valle L.~F.,   Kacprzak T.,  2024, \mn@doi [Journal of Cosmology and Astroparticle Physics] {10.1088/1475-7516/2024/04/023}, 2024, 023

\bibitem[\protect\citeauthoryear{Bobkov \& Ledoux}{Bobkov \& Ledoux}{2019}]{bobkov2019kantorovich}
Bobkov S.,  Ledoux M.,  2019, One-dimensional empirical measures, order statistics, and Kantorovich transport distances.
 Memoirs of the American Mathematical Society Vol. 261, American Mathematical Society

\bibitem[\protect\citeauthoryear{Brenier, Frisch, H{\'e}non, Loeper, Matarrese, Mohayaee  \& Sobolevski{\u\i}}{Brenier et~al.}{2003}]{brenier2003reconstruction}
Brenier Y.,  Frisch U.,  H{\'e}non M.,  Loeper G.,  Matarrese S.,  Mohayaee R.,   Sobolevski{\u\i} A.,  2003, Monthly Notices of the Royal Astronomical Society, 346, 501

\bibitem[\protect\citeauthoryear{{Chaves-Montero}, Angulo, Schaye, Schaller, Crain, Furlong  \& Theuns}{{Chaves-Montero} et~al.}{2016}]{chaves-monteroSubhaloAbundanceMatching2016}
{Chaves-Montero} J.,  Angulo R.~E.,  Schaye J.,  Schaller M.,  Crain R.~A.,  Furlong M.,   Theuns T.,  2016, \mn@doi [Monthly Notices of the Royal Astronomical Society] {10.1093/mnras/stw1225}, 460, 3100

\bibitem[\protect\citeauthoryear{Cominetti \& Mart{\'\i}n}{Cominetti \& Mart{\'\i}n}{1994}]{cominetti1994asymptotic}
Cominetti R.,  Mart{\'\i}n J.~S.,  1994, Mathematical Programming, 67, 169

\bibitem[\protect\citeauthoryear{Conroy, Wechsler  \& Kravtsov}{Conroy et~al.}{2006}]{conroyModelingLuminositydependentGalaxy2006}
Conroy C.,  Wechsler R.~H.,   Kravtsov A.~V.,  2006, \mn@doi [The Astrophysical Journal] {10.1086/503602}, 647, 201

\bibitem[\protect\citeauthoryear{Cuturi}{Cuturi}{2013}]{cuturi2013sinkhorn}
Cuturi M.,  2013, Advances in neural information processing systems, 26

\bibitem[\protect\citeauthoryear{Dragomir, {Rodr{\'i}guez-Puebla}, Primack  \& Lee}{Dragomir et~al.}{2018}]{dragomirDoesGalaxyHalo2018}
Dragomir R.,  {Rodr{\'i}guez-Puebla} A.,  Primack J.~R.,   Lee C.~T.,  2018, \mn@doi [Monthly Notices of the Royal Astronomical Society] {10.1093/mnras/sty283}, 476, 741

\bibitem[\protect\citeauthoryear{Fischbacher, Kacprzak, Blazek  \& Refregier}{Fischbacher et~al.}{2023}]{fischbacherRedshiftRequirementsCosmic2023}
Fischbacher S.,  Kacprzak T.,  Blazek J.,   Refregier A.,  2023, \mn@doi [Journal of Cosmology and Astroparticle Physics] {10.1088/1475-7516/2023/01/033}, 2023, 033

\bibitem[\protect\citeauthoryear{Fischbacher, Moser, Kacprzak, Herbel, Tortorelli, Schmitt, Refregier  \& Amara}{Fischbacher et~al.}{2024}]{fischbachergalsbiPythonPackage2024}
Fischbacher S.,  Moser B.,  Kacprzak T.,  Herbel J.,  Tortorelli L.,  Schmitt U.,  Refregier A.,   Amara A.,  2024, $\texttt{galsbi}$: {{A Python}} Package for the {{GalSBI}} Galaxy Population Model (\mn@eprint {arXiv} {2412.08722}), \mn@doi{10.48550/arXiv.2412.08722}

\bibitem[\protect\citeauthoryear{Fischbacher, Kacprzak, Tortorelli, Moser, Refregier, Gebhardt  \& Gruen}{Fischbacher et~al.}{2025}]{fischbacherGalSBIPhenomenologicalGalaxy2024}
Fischbacher S.,  Kacprzak T.,  Tortorelli L.,  Moser B.,  Refregier A.,  Gebhardt P.,   Gruen D.,  2025, \mn@doi [Journal of Cosmology and Astroparticle Physics] {10.1088/1475-7516/2025/06/007}, 2025, 007

\bibitem[\protect\citeauthoryear{Flamary et~al.,}{Flamary et~al.}{2021}]{flamaryPOTPythonOptimal2021}
Flamary R.,  et~al., 2021, Journal of Machine Learning Research, 22, 1

\bibitem[\protect\citeauthoryear{Guo, White, Li  \& {Boylan-Kolchin}}{Guo et~al.}{2010}]{guoHowGalaxiesPopulate2010}
Guo Q.,  White S.,  Li C.,   {Boylan-Kolchin} M.,  2010, \mn@doi [Monthly Notices of the Royal Astronomical Society] {10.1111/j.1365-2966.2010.16341.x}, 404, 1111

\bibitem[\protect\citeauthoryear{Hitz, Berner, Crichton, Hennig  \& Refregier}{Hitz et~al.}{2025}]{hitzFastSimulationCosmological2024}
Hitz P.,  Berner P.,  Crichton D.,  Hennig J.,   Refregier A.,  2025, \mn@doi [Journal of Cosmology and Astroparticle Physics] {10.1088/1475-7516/2025/04/003}, 2025, 003

\bibitem[\protect\citeauthoryear{Hunter}{Hunter}{2007}]{hunterMatplotlib2DGraphics2007}
Hunter J.~D.,  2007, \mn@doi [Computing in Science Engineering] {10.1109/MCSE.2007.55}, 9, 90

\bibitem[\protect\citeauthoryear{Jones, Haynes, Giovanelli  \& Moorman}{Jones et~al.}{2018}]{jones2018alfalfa}
Jones M.~G.,  Haynes M.~P.,  Giovanelli R.,   Moorman C.,  2018, Monthly Notices of the Royal Astronomical Society, 477, 2

\bibitem[\protect\citeauthoryear{Jordan, Kinderlehrer  \& Otto}{Jordan et~al.}{1998}]{jordan1998fokkerplanck}
Jordan R.,  Kinderlehrer D.,   Otto F.,  1998, SIAM journal on mathematical analysis, 29, 1

\bibitem[\protect\citeauthoryear{Kacprzak \& Fluri}{Kacprzak \& Fluri}{2022}]{kacprzakDeepLSSBreakingParameter2022}
Kacprzak T.,  Fluri J.,  2022, \mn@doi [Physical Review X] {10.1103/PhysRevX.12.031029}, 12, 031029

\bibitem[\protect\citeauthoryear{Kantorovich}{Kantorovich}{1942}]{kantorovich1942translocation}
Kantorovich L.~V.,  1942, in Dokl. Akad. Nauk. USSR (NS). pp 199--201

\bibitem[\protect\citeauthoryear{Kravtsov, Berlind, Wechsler, Klypin, Gottlober, Allgood  \& Primack}{Kravtsov et~al.}{2004}]{kravtsovDarkSideHalo2004}
Kravtsov A.~V.,  Berlind A.~A.,  Wechsler R.~H.,  Klypin A.~A.,  Gottlober S.,  Allgood B.,   Primack J.~R.,  2004, \mn@doi [The Astrophysical Journal] {10.1086/420959}, 609, 35

\bibitem[\protect\citeauthoryear{Masaki, Lin  \& Yoshida}{Masaki et~al.}{2013}]{Masaki_Lin_Yoshida_2013}
Masaki S.,  Lin Y.-T.,   Yoshida N.,  2013, \mn@doi [Monthly Notices of the Royal Astronomical Society] {10.1093/mnras/stt1729}, 436, 2286–2300

\bibitem[\protect\citeauthoryear{Mo, {van den Bosch}  \& White}{Mo et~al.}{2010}]{moGalaxyFormationEvolution2010}
Mo H.,  {van den Bosch} F.~C.,   White S.,  2010, Galaxy {{Formation}} and {{Evolution}}

\bibitem[\protect\citeauthoryear{Monaco, Theuns  \& Taffoni}{Monaco et~al.}{2002a}]{monacoPINOCCHIOPinpointingOrbitcrossing2002}
Monaco P.,  Theuns T.,   Taffoni G.,  2002a, \mn@doi [Monthly Notices of the Royal Astronomical Society] {10.1046/j.1365-8711.2002.05162.x}, 331, 587

\bibitem[\protect\citeauthoryear{Monaco, Theuns, Taffoni, Governato, Quinn  \& Stadel}{Monaco et~al.}{2002b}]{monacoPredictingNumberSpatial2002}
Monaco P.,  Theuns T.,  Taffoni G.,  Governato F.,  Quinn T.,   Stadel J.,  2002b, \mn@doi [The Astrophysical Journal] {10.1086/324182}, 564, 8

\bibitem[\protect\citeauthoryear{Monaco, Sefusatti, Borgani, Crocce, Fosalba, Sheth  \& Theuns}{Monaco et~al.}{2013}]{monacoAccurateToolFast2013}
Monaco P.,  Sefusatti E.,  Borgani S.,  Crocce M.,  Fosalba P.,  Sheth R.~K.,   Theuns T.,  2013, \mn@doi [Monthly Notices of the Royal Astronomical Society] {10.1093/mnras/stt907}, 433, 2389

\bibitem[\protect\citeauthoryear{Moser, Lorenz, Schmitt, R{\'e}fr{\'e}gier, Fluri, Sgier, Tarsitano  \& Heisenberg}{Moser et~al.}{2022}]{moserSymbolicImplementationExtensions2022}
Moser B.,  Lorenz C.~S.,  Schmitt U.,  R{\'e}fr{\'e}gier A.,  Fluri J.,  Sgier R.,  Tarsitano F.,   Heisenberg L.,  2022, \mn@doi [Astronomy and Computing] {10.1016/j.ascom.2022.100603}, 40, 100603

\bibitem[\protect\citeauthoryear{Moster, Somerville, Maulbetsch, {van den Bosch}, Macci{\`o}, Naab  \& Oser}{Moster et~al.}{2010}]{mosterConstraintsRelationshipStellar2010}
Moster B.~P.,  Somerville R.~S.,  Maulbetsch C.,  {van den Bosch} F.~C.,  Macci{\`o} A.~V.,  Naab T.,   Oser L.,  2010, \mn@doi [The Astrophysical Journal] {10.1088/0004-637X/710/2/903}, 710, 903

\bibitem[\protect\citeauthoryear{Munari, Monaco, Sefusatti, Castorina, Mohammad, Anselmi  \& Borgani}{Munari et~al.}{2017}]{munariImprovingFastGeneration2017}
Munari E.,  Monaco P.,  Sefusatti E.,  Castorina E.,  Mohammad F.~G.,  Anselmi S.,   Borgani S.,  2017, \mn@doi [Monthly Notices of the Royal Astronomical Society] {10.1093/mnras/stw3085}, 465, 4658

\bibitem[\protect\citeauthoryear{Nutz}{Nutz}{2024}]{nutz2024quadratically}
Nutz M.,  2024, arXiv preprint arXiv:2404.06847

\bibitem[\protect\citeauthoryear{Nuza et~al.,}{Nuza et~al.}{2013}]{nuzaClusteringGalaxies052013}
Nuza S.~E.,  et~al., 2013, \mn@doi [Monthly Notices of the Royal Astronomical Society] {10.1093/mnras/stt513}, 432, 743

\bibitem[\protect\citeauthoryear{Pass}{Pass}{2015}]{pass2015multimargot}
Pass B.,  2015, ESAIM: Mathematical Modelling and Numerical Analysis, 49, 1771

\bibitem[\protect\citeauthoryear{Reddick, Wechsler, Tinker  \& Behroozi}{Reddick et~al.}{2013}]{reddickCONNECTIONGALAXIESDARK2013}
Reddick R.~M.,  Wechsler R.~H.,  Tinker J.~L.,   Behroozi P.~S.,  2013, \mn@doi [The Astrophysical Journal] {10.1088/0004-637X/771/1/30}, 771, 30

\bibitem[\protect\citeauthoryear{Refregier, Gamper, Amara  \& Heisenberg}{Refregier et~al.}{2018}]{refregierPyCosmoIntegratedCosmological2018}
Refregier A.,  Gamper L.,  Amara A.,   Heisenberg L.,  2018, \mn@doi [Astronomy and Computing] {10.1016/j.ascom.2018.08.001}, 25, 38

\bibitem[\protect\citeauthoryear{Rizzo, {Villaescusa-Navarro}, Monaco, Munari, Borgani, Castorina  \& Sefusatti}{Rizzo et~al.}{2017}]{rizzoSimulatingCosmologies$L$CDM2017}
Rizzo L.~A.,  {Villaescusa-Navarro} F.,  Monaco P.,  Munari E.,  Borgani S.,  Castorina E.,   Sefusatti E.,  2017, \mn@doi [Journal of Cosmology and Astroparticle Physics] {10.1088/1475-7516/2017/01/008}, 2017, 008

\bibitem[\protect\citeauthoryear{{Rodr{\'i}guez-Puebla}, {Avila-Reese}, Yang, Foucaud, Drory  \& Jing}{{Rodr{\'i}guez-Puebla} et~al.}{2015}]{rodriguez-pueblaStelLARTOHALOMASSRELATION2015}
{Rodr{\'i}guez-Puebla} A.,  {Avila-Reese} V.,  Yang X.,  Foucaud S.,  Drory N.,   Jing Y.~P.,  2015, \mn@doi [The Astrophysical Journal] {10.1088/0004-637X/799/2/130}, 799, 130

\bibitem[\protect\citeauthoryear{Simha, Weinberg, Dave, Fardal, Katz  \& Oppenheimer}{Simha et~al.}{2012}]{simhaTestingSubhaloAbundance2012}
Simha V.,  Weinberg D.,  Dave R.,  Fardal M.,  Katz N.,   Oppenheimer B.~D.,  2012, \mn@doi [Monthly Notices of the Royal Astronomical Society] {10.1111/j.1365-2966.2012.21142.x}, 423, 3458

\bibitem[\protect\citeauthoryear{Sinkhorn}{Sinkhorn}{1964}]{sinkhorn1964relationship}
Sinkhorn R.,  1964, Annals of Mathematical Statistics, 35, 876

\bibitem[\protect\citeauthoryear{Taffoni, Monaco  \& Theuns}{Taffoni et~al.}{2002}]{taffoniPINOCCHIOHierarchicalBuildup2002}
Taffoni G.,  Monaco P.,   Theuns T.,  2002, \mn@doi [Monthly Notices of the Royal Astronomical Society] {10.1046/j.1365-8711.2002.05441.x}, 333, 623

\bibitem[\protect\citeauthoryear{Tarsitano et~al.,}{Tarsitano et~al.}{2021}]{tarsitanoPredictingCosmologicalObservables2021}
Tarsitano F.,  et~al., 2021, \mn@doi [Astronomy and Computing] {10.1016/j.ascom.2021.100484}, 36, 100484

\bibitem[\protect\citeauthoryear{{Trujillo-Gomez}, Klypin, Primack  \& Romanowsky}{{Trujillo-Gomez} et~al.}{2011}]{trujillo-gomezGALAXIESLCDMHALO2011}
{Trujillo-Gomez} S.,  Klypin A.,  Primack J.,   Romanowsky A.~J.,  2011, \mn@doi [The Astrophysical Journal] {10.1088/0004-637X/742/1/16}, 742, 16

\bibitem[\protect\citeauthoryear{Vale \& Ostriker}{Vale \& Ostriker}{2004}]{valeLinkingHaloMass2004}
Vale A.,  Ostriker J.~P.,  2004, \mn@doi [Monthly Notices of the Royal Astronomical Society] {10.1111/j.1365-2966.2004.08059.x}, 353, 189

\bibitem[\protect\citeauthoryear{Van Der~Walt, Colbert  \& Varoquaux}{Van Der~Walt et~al.}{2011}]{vanderwaltNumPyArrayStructure2011}
Van Der~Walt S.,  Colbert S.~C.,   Varoquaux G.,  2011, \mn@doi [Computing in Science \& Engineering] {10.1109/MCSE.2011.37}, 13, 22

\bibitem[\protect\citeauthoryear{Villani}{Villani}{2003}]{villani2021ot}
Villani C.,  2003, Topics in Optimal Transportation.
 Graduate Studies in Mathematics Vol. 58, American Mathematical Society

\bibitem[\protect\citeauthoryear{Villani}{Villani}{2009}]{villani2009otoldnew}
Villani C.,  2009, Optimal Transport: Old and New, 1 edn.
Grundlehren der mathematischen Wissenschaften, Springer Berlin, Heidelberg, \mn@doi{10.1007/978-3-540-71050-9}

\bibitem[\protect\citeauthoryear{Virtanen et~al.,}{Virtanen et~al.}{2020}]{virtanenSciPy10FundamentalAlgorithms2020}
Virtanen P.,  et~al., 2020, \mn@doi [Nature Methods] {10.1038/s41592-019-0686-2}, 17, 261

\bibitem[\protect\citeauthoryear{Watson, Berlind  \& Zentner}{Watson et~al.}{2012}]{watsonConstrainingSatelliteGalaxy2012}
Watson D.~F.,  Berlind A.~A.,   Zentner A.~R.,  2012, \mn@doi [The Astrophysical Journal] {10.1088/0004-637X/754/2/90}, 754, 90

\bibitem[\protect\citeauthoryear{Watson, Iliev, D'Aloisio, Knebe, Shapiro  \& Yepes}{Watson et~al.}{2013}]{watsonHaloMassFunction2013}
Watson W.~A.,  Iliev I.~T.,  D'Aloisio A.,  Knebe A.,  Shapiro P.~R.,   Yepes G.,  2013, \mn@doi [Monthly Notices of the Royal Astronomical Society] {10.1093/mnras/stt791}, 433, 1230

\bibitem[\protect\citeauthoryear{Wechsler \& Tinker}{Wechsler \& Tinker}{2018}]{wechslerConnectionGalaxiesTheir2018}
Wechsler R.~H.,  Tinker J.~L.,  2018, \mn@doi [Annual Review of Astronomy and Astrophysics] {10.1146/annurev-astro-081817-051756}, 56, 435

\bibitem[\protect\citeauthoryear{Wetzel \& White}{Wetzel \& White}{2010}]{wetzelWhatDeterminesSatellite2010}
Wetzel A.~R.,  White M.,  2010, \mn@doi [Monthly Notices of the Royal Astronomical Society] {10.1111/j.1365-2966.2009.16191.x}, 403, 1072

\bibitem[\protect\citeauthoryear{Yamamoto, Masaki  \& Hikage}{Yamamoto et~al.}{2015}]{yamamotoTestingSubhaloAbundance2015}
Yamamoto M.,  Masaki S.,   Hikage C.,  2015, Testing Subhalo Abundance Matching from Redshift-Space Clustering (\mn@eprint {arXiv} {1503.03973}), \mn@doi{10.48550/arXiv.1503.03973}

\bibitem[\protect\citeauthoryear{Zentner, Hearin  \& van~den Bosch}{Zentner et~al.}{2014}]{zentnerGalaxyAssemblyBias2014}
Zentner A.~R.,  Hearin A.~P.,   van~den Bosch F.~C.,  2014, \mn@doi [Monthly Notices of the Royal Astronomical Society] {10.1093/mnras/stu1383}, 443, 3044

\makeatother
\end{thebibliography}

%%%%%%%%%%%%%%%%%%%%%%%%%%%%%%%%%%%%%%%%%%%%%%%%%%

% Don't change these lines
\bsp	% typesetting comment
\label{lastpage}
\end{document}